\documentclass[twocolumn,showpacs,preprintnumbers,amsmath,amssymb]{revtex4}
\usepackage{graphicx}% Include figure files
\usepackage{dcolumn}% Align table columns on decimal point
\usepackage{bm}

\newcommand{\eqb}{\begin{equation}}
\newcommand{\eqe}{\end{equation}}

\newcommand{\eab}{\begin{eqnarray}}
\newcommand{\eae}{\end{eqnarray}}

\begin{document}

\title{Pointlike electrons and muons and 
the nature of neutrinos}
\author{Ralf Hofmann}\vspace{0.3cm}
\affiliation{Institut f\"ur Theoretische Physik,
Universit\"at Heidelberg,
Philosophenweg 16,
69120~Heidelberg,
Germany}

\begin{abstract}

We show that the form factor of an electron or a muon 
is unity for the modulus of the momentum transfer away 
from the Hagedorn temperatures $T^{e}_0\sim m_e$ and $T^{\mu}_0\sim m_\mu$ 
and lower than the hadronic Hagedorn temperature $T^{\tiny\mbox{QCD}}_0\sim m_\pi$. 
The structure of the electron or the 
muon is resolved for momemtum transfers comparable to $T^{e,\mu}_0$. This is seen in 
scattering experiments performed in the 1950s, 1960s, and early 1970s to test the validity of QED.  
We interprete center-vortex loops with no self-intersection, occurring in 
the confining phases of each SU(2) gauge-theory factor, as {\sl Majorana} neutrinos. 
We offer a possible explanation for a contribution to 
$\Lambda_{\tiny\mbox{cosmo}}$ not generated by the so-far nonconfining, 
strongly interacting gauge theory SU(2)$_{\tiny{CMB}}$. 


\end{abstract} 
%$\mbox{}$\\ 

\maketitle
%*********************

\indent {\sl Introduction.} 
The main purpose of this Letter is to extend and substantialize a thermal 
theory for charged leptons put forward 
in \cite{Hofmann20032}. The work in \cite{Hofmann20032} is based on the approach in 
\cite{Hofmann20031}. The 
{\sl extension} is related to the {\sl existence} and the {\sl nature} of neutrinos. 
We will not address the {\sl interactions} of neutrinos with charged leptons. 
In the present Universe the gauge dynamics subject to the symmetry 
SU(2)$_{\tiny\mbox{CMB}}\times$
SU(2)$_{e}\times$SU(2)$_{\mu}\times$SU(2)$_{\tau}$ predicts the existence 
of stable fermionic \cite{Hofmann20033} states (charged leptons) with one unit of electric 
charge \cite{Hofmann20032}. In addition, it predicts the existence of one (entangled) 
asymptotic photon. The electric charge is a magnetic charge in each of the SU(2) 
gauge-theory factors. Its coupling to the photon (the fine-structure constant $\alpha$) can be 
postdicted to 2.4\% accuracy in a tree-level treatment of the thermodynamics. 
Our theory seems to resolve the long-standing 
problem of the apparent infinite self-energy of a structureless 
charged lepton and most probably a number 
of other important problems \cite{Hofmann20032}. We believe that the gauge theory with symmetry 
SU(2)$_{\tiny\mbox{CMB}}\times$
SU(2)$_{e}\times$SU(2)$_{\mu}\times$SU(2)$_{\tau}$ is underlying Quantum Electrodynamics (QED).
Here we {\sl substantialize} this theory in view of the following apparent contradiction: 
In scattering, annihilation, and production processes involving purely 
electrodynamical interactions one would expect to see the structure of 
the charged lepton unless some so-far unknown mechanism prevents 
Nature from showing this structure to us most of the time. As we will argue below, 
this mechanism is rooted in the existence of an 
exponentially growing density of extremely instable excitations.\\  
\indent {\sl Mini-review of experimental situation.} The elastic differential cross 
section for $e^-e^-$ ($e^-e^+$) scattering was measured in \cite{Ashkin1953} 
for electron (positron) incident kinetic energies 
between $0.6\,\mbox{MeV}$ and $1.2\,\mbox{MeV}$ ($0.6\,\mbox{MeV}$ and $1.0\,\mbox{MeV}$) 
and an energy transfer of 50\% the incident kinetic energy. For Moeller (Bhabha) scattering good 
agreement of experiment with theory was observed at the high-energy 
ends. At $0.6\,\mbox{MeV}$ incident kinetic energy the differential cross sections 
turned out to be significantly lower than the 
theoretical Moeller (Bhabha) prediction (it is clear that 
binding effects are entirely negligible). 
Measurements of $e^-e^-$ elastic 
scattering at 6.1\,MeV and 15.7\,MeV incident kinetic 
electron energy were reported in \cite{Barber1953,Scott1951}. Excellent agreement with the 
Moeller prediction was obtained. In the 1960s and early 1970s 
violations of (tree-level) QED were parameterized  
by a cutoff scale in the form factor 
$F_\Lambda(q^2)=(1-q^2/\Lambda^2)^{-1}$ of the charged lepton. 
Elastic $e^-e^-$ scattering experiments gave different values 
for $\Lambda$: $\Lambda>0.76\,$GeV (4.4\,GeV) for 
$e^-e^-$ scattering at $\sqrt{s}=600\,$MeV ($\sqrt{s}=1112\,$MeV) 
and 95\% confidence \cite{Barber1966} (\cite{Barber1971}). A test of 
QED by $e^-e^+$ scattering at $\sqrt{s}=2\times$ 510 MeV yielded $\Lambda>3.8\,$GeV (positive metric) and 
$\Lambda>2.0\,$GeV (negative metric) \cite{Augustin1970}. Experiments using a 
bremsstrahlungsbeam incident on a carbon target to produce $e^+e^-$ 
pairs at wide angles \cite{Blumenthal1966,Alvensleben1968} obtain 
the ratio $R$ of experimental, differential cross section to the 
Bethe-Heitler prediction (the contribution of Compton processes is very small). 
In \cite{Blumenthal1966} $R$ was found to deviate strongly from 
unity for $e^+e^-$ invariant masses $M$ in the vicinity of the muon threshold. 
In particular, for $M<170\,$MeV $R$ was measured to be significantly 
below unity whereas for $M>200\,$MeV $R$ values were measured consistent 
with or larger than unity. In \cite{Alvensleben1968} the experiment 
was repeated, and a much better consistency of $R$ 
with unity was claimed. Still, $R$ was measured to 
be significantly below unity at $M\sim 170,310,450$\,MeV and above unity at $M\sim 380\,$MeV. 
For $500<M<900$\,MeV experiment turned out to be consistent with $R=1$. 
Similar results were obtained in
\cite{Asbury1967}. In \cite{Balakin19711} a rather weak 
lower bound of $\Lambda>1.3\,$GeV was measured 
in $e^-e^+$ annihilation into 2$\gamma$ at $\sqrt{s}=1\,$GeV, and 
in \cite{Balakin19712} the QED tree-level prediction for total muon production cross section by 
$e^+e^-\to\mu^+\mu^-$ was found to be consistent with experiment at $1020\,\mbox{MeV}<\sqrt{s}
<1340\,\mbox{MeV}$. In \cite{Bernadini1971} data on wide-angle $e^+e^-$ Bhabha scattering at 
$1.4\,\mbox{GeV}<\sqrt{q^2}<2.4\,\mbox{GeV}$ are reported. In the vicinity of the $\tau$ 
threshold significant deviations of the measured cross section from the QED prediction were 
seen (three data points at $\sqrt{q^2}=1.8,1.85,2.4$ were not consistent with the 
QED prediction). For a compilation of QED tests at higher energy see \cite{DittmannHepp1981}. 
A conspicuous feature is that differential cross sections for the reaction 
$e^+e^-\to e^+e^-$ deviate significantly from the QED prediction at small scattering 
angle where the momentum transfer is close to the kinematic threshold for 
$\tau$ production. To conclude, the data hint on 
deviations of experimental cross sections from the QED prediction in 
the vicinity of the kinematical threshold for the production of a charged lepton. 
Away from 
this threshold (tree-level) QED passes the experimental 
tests very well. But why does tree-level QED give such a good description 
in the latter case even though we believe now that charged 
leptons are solitons? Before we give a partial 
answer to this question (partial because we so far have no handle on QCD effects, see below) 
we would like to remind the reader of the fact that 
in the real world the above gauge dynamics is  `contaminated' 
by the hadronic world subject to the chromodynamical 
gauge symmetry SU(3)$_C$. This effect becomes important 
as the momentum transfer felt by a charged lepton exceeds the 
pion production threshold $m_\pi\sim 140\,$MeV. 
Since we would like to discuss pure QED and (noninteracting) neutrinos 
we restrict ourselves to a discussion of the gauge dynamics subject to 
SU(2)$_{\tiny\mbox{CMB}}\times$SU(2)$_{e}\times$SU(2)$_{\mu}$ and 
to momentum transfers smaller than $m_\pi$. An inclusion of 
QCD effects in our thermodynamical approach would need a genuine 
understanding of the nature of quarks (maybe quarks are 
color magnetic and magnetic monopoles of a pure SU(N$>$3) gauge theory?). 
This lack of our present understanding manifests itself in the 
unexpected discoveries of narrow hadronic states 
with more than three quarks as 
constituents \cite{pentaquark}). \\ 
\indent {\sl Hagedorn's approach to particle collisions in strongly interacting theories.} 
Our discussion of the apparent structurelessness of the electron and the muon 
heavily relies on Hagedorn's work on the thermodynamics of strong interactions 
\cite{Hagedorn1965} and the `mechanism independent of the method of excitation' 
by Wu and Yang \cite{WuYang1964}. Therefore, we briefly review 
the main ideas and results. Hagedorn 
bases his approach on the following three postulates:\vspace{.1cm}\\ 
(i) strong interactions are so strong that they produce an infinity of resonances 
(for resonance mass $\to\infty$ they are called fire-balls), \\ 
(ii) fire-balls can be described by statistical thermodynamics,\\  
(iii) fire-balls consist of fire-balls.\vspace{.1cm}\\ 
In addition:\vspace{.1cm}\\ 
(i$^\prime$) strong interaction are as strong as 
they possibly can be without violating postulate (ii).\vspace{.1cm}\\ 
Postulates (i) and (ii) enable Hagedorn to define partition functions 
for the strongly interacting system (i) in terms 
of hadronic resonances and (ii) in terms of fire-balls. They are dual descriptions 
of one and the same physics which involve densities of states $\rho(m)$ and $\sigma(E)$, 
respectively. A weak condition of self-consistency is the equality 
of the entropies at large energies
%************
\eab
\label{scc}
\log{\rho(x)}&\to& \log{\sigma(x)}\,,\ \ \ \ x\to\infty\,.
\eae
%*********** 
Hagedorn then argues that the only admissible solutions to this condition are exponentially 
increasing densities of state. The occurrence of a highest 
temperature $T_0$ is an immediate consequence of this solution. 
Subsequently, Hagedorn applies his theory to a description 
of scattering and annihilation processes in strong interactions. For center-of-mass 
energies considerably (but not too much) larger than the highest temperature 
$T_0$ there is no real difference between bosonic and fermionic statistics. The 
probability density $w(p_\perp)$ for generating a particle 
of mass $m$ and transverse 
momentum $p_\perp$ in the collision is derived as
%***********       
\eab
\label{prop}
w(p_\perp)&=&\mbox{const}\times p_\perp
\sqrt{T_0}\sqrt{p_\perp^2}+\nonumber\\ 
& &m^2\exp\left[-1/T_0\sqrt{p_\perp^2+m^2}\right]\,.
\eae
%***********
For example, a differential elastic cross 
section for $2\to 2$ scattering contains three types of factors: (a) $w(p_\perp)$ in (\ref{prop}), 
(b) the probability that the number of particles is two, and (c) 
kinematical and geometrical factors being algebraic in 
$p_\perp$ and $E$. Wu and Yang observe \cite{WuYang1964} 
that the {\sl electromagnetic} form factor $F(q^2)$ of the 
proton (we do not distinguish between electric and magnetic form factors in this Letter) 
is an exponentially falling function of the momentum transfer $\sqrt{q^2}$ as
%*********
\eab
\label{WuYang}
F(q^2)\to \exp\left[-\sqrt{q^2}/(4T_0)\right]\,\ \ \ \ (\sqrt{q^2}\gg T_0)\,.
\eae
%**********
Hagedorn interpretes Eq.\,(\ref{WuYang}) as a supporting fact 
for his thermodynamical theory for particle scattering in strong interactions. 
He obtains an estimate for $T_0\sim 165\,$MeV by taking the inverse pion mass 
as a natural radius for the fire-ball mediating the scattering of hadrons. \\ 
\indent {\sl QED, an exponentially rising density of states, and neutrinos.} Hagedorn developed 
his thermodynamical approach to particle scattering within the framework of strong 
interactions. He did not anticipate the possiblity that Quantum Electrodynamics 
(QED) could be generated by a strongly interacting theory in the sense proposed in 
\cite{Hofmann20032}. This theory {\sl possesses} an exponentially growing density of states. 
The spectrum of states in the confining (or center) phase for one of the SU(2) gauge 
theories in SU(2)$_{\tiny\mbox{CMB}}\times$
SU(2)$_{e}\times$SU(2)$_{\mu}\times$SU(2)$_{\tau}$ is 
depicted in Fig.\,1. 
In \cite{Hofmann20032} we have interpreted the single center-vortex 
loop (the first state in Fig.\,1) 
%***********************
\begin{figure}
\begin{center}
\leavevmode
\leavevmode
%\epsffile[80 25 534 344]{}
\vspace{3.5cm}
\includegraphics{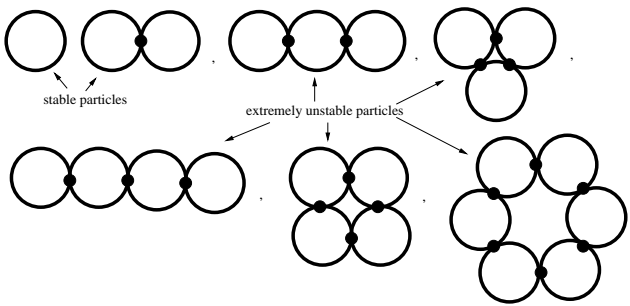}
\end{center}
\caption{The spectrum of excitations of an SU(2) gauge theory in 
its confining phase. The thick lines denote (large) center-vortex loops, 
and the dots represent crossings of center vortices. Each crossing carries 
$\pm$ one unit of electric charge. The first two states
correspond to stable excitations. 
States with more than one center-vortex crossing 
are extremely unstable in the presence of photons which belong to 
SU(2) gauge-theory factors not being in their center (or confining) phases. 
The electric charges in such `hadrons' either 
(very rarely) annihilate into neutrinos \cite{neutrinos} or (much more often) into 
photons. A `hadron' is torn apart by Coulomb 
repulsion if only equal-sign electric charges reside in it.}      
\end{figure}
%************************
as a dark matter particle since it does not carry electric charge. We did not specify further 
the nature of this particle. At first sight we have rejected 
the possibility that this particle is the 
neutrino associated with the charged lepton. This seemed to be justified by the (erroneous) 
conclusion that the mass of the single center-vortex 
loop should be comparable to the mass of the charged lepton. 
In the meantime we have changed our conclusion. A center-vortex loop with no self-intersection has vanishing 
energy in the confining phase for a sufficiently large loop size 
\cite{LangeEngelhardtReinhardt2003}. This is the reason why center-vortices 
condense in the first place (in \cite{BruckmannEngelhardt2003} 
a beautiful discussion is given of the 
distribution of topological charge in dependence of the 
gradient of the tangential vector to the loop). Thus, the associated particle has vanishing 
mass. There is an experimental upper bound on the mass of the 
electron-neutrino ($\nu_e$) of 2.2\,eV \cite{Mainz} and 2.5\,eV \cite{Troitsk}. 
Compared to the electron mass $m_e=511$\,keV this is a
tiny number. In the Troitsk experiment \cite{Troitsk} a final 
bump in the Kurie plot was reported. This bump led the authors of 
\cite{BosiCavalleri2002} to the 
interpretion of $\nu_e$ as a zero-point 
fluctuation. This conclusion is quite natural if the neutrino is interpreted as a 
large center-vortex loop. A further conclusion is 
that the tiny neutrino masses follow the same hierarchical mass 
pattern as the charged leptons do since 
the tiny energy of a single center-vortex loop 
is governed by the size of it (same for all families) and the 
associated Yang-Mills scale 
or the Hagedorn temperature or the mass of the charged lepton. 
As for the spin, it was found in \cite{Hofmann20033} 
that each stable particle in the center (or confining) phase of 
an SU(2) gauge theory should be 
interpreted as a fermion. Assuming thermalization sufficiently 
long after the 1$^{\tiny\mbox{st}}$ order phase transition 
towards the center phase, 
this property is derived from counting inequivalent, 
nonlocal $Z_2$ transformations which change the periodic boundary conditions of the 
underlying SU(2) Yang-Mills theory into 
antiperiodic ones during the phase transition. 
Since there is no way of distinguishing a neutrino from 
its antiparticle in our theory we are led to conclude 
that neutrinos are of the {\sl Majorana} type \cite{Majoranaexp}. This 
conclusion is supported experimentally by the evidence for double-beta 
decay of $\mbox{}^{76}$Ge as published in \cite{Klein-Groothaus}.       

The gauge group describing electrodynamics was proposed in \cite{Hofmann20031} 
to be SU(2)$_{\tiny\mbox{CMB}}\times$
SU(2)$_{e}\times$SU(2)$_{\mu}\times$SU(2)$_{\tau}$. For momentum transfers $\sqrt{q^2}$ 
much smaller than the Hagedorn temperature $T_0^e$ but larger than 
$T_0^{\tiny\mbox{CMB}}\sim 10^{-4}$\,eV there is only 
one massless photon in the theory \cite{Hofmann20032}. This photon is quantum 
entangled with two very heavy photons which belong to the two 
gauge-theory factors in SU(2)$_e\times$SU(2)$_\mu$ being in 
their confining phases \cite{Hofmann20032}. In an on-shell state 
(no restriction on its energy) the massless photon can be considered as a part of 
the thermal ensemble, characterized by the
temperature $T_{\tiny\mbox{CMB}}$. This is true because the 
photon, possibly a radio wave or visible light or a $\gamma$ ray, 
only generates a temperature 
fluctuation $\Delta T/T_{\tiny\mbox{CMB}}\ll 1$ which 
is much smaller than the primordial fluctuations 
seen in the CMB. We thus may view the entire visible Universe as a 
fire-ball of temperature $T_{\tiny\mbox{CMB}}$. The mechanism which 
decreases the temperature of this fire-ball is gravitational expansion. 
For elastic scattering of charged leptons the photon couples to the respective lepton 
via a form factor which is written as a 
sum (recall postulate (iii) above, the scattering is a statistical process 
involving {\sl all} purely leptonic resonances) of the type
%*********
\eab
\label{FF}
F(q^2)&\sim& 1/2\left(\exp\left[-\sqrt{q^2}/(4T^e_0)\right]+
\exp\left[-\sqrt{q^2}/(4T^\mu_0)\right]\right)\,,\nonumber\\ & &
\eae
%**********
where $T^e_0$ and $T^\mu_0$ denote the Hagedorn 
temperatures belonging to the factors in 
SU(2)$_{e}\times$SU(2)$_{\mu}$. We have omitted the form 
factor due to $T^{\tiny\mbox{CMB}}_0$ since 
for any so-far performed QED elastic scattering, annihilation or production 
experiment the momentum transfers are much larger than this 
temperature, and thus the exponential 
suppression is extremely large. The $\sim$ sign in (\ref{FF}) 
indicates that the exponential 
dependence is only valid sufficiently far above {\sl or} 
below a kinematical threshold for the production of a charged 
lepton (in practice a factor of two or three is sufficient). Close to the 
threshold the exponential should be
replaced by a Hankel function \cite{Hagedorn1965}. The normalization 
$1/2$ defines the fine-structure constant 
to be $\alpha=1/137$ for momentum transfer considerably smaller 
than $T^e_0$ \cite{Hofmann20032}.  
%***********************
\begin{figure}
\begin{center}
\leavevmode
\leavevmode
%\epsffile[80 25 534 344]{}
\vspace{3.5cm}
\includegraphics{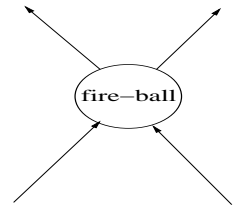}
\end{center}
\caption{Elastic scattering of charged leptons.}      
\end{figure}
%************************
As for the $\tau$ lepton, 
its mass $m_\tau\sim 1777\,$MeV is much larger 
than the hadronic threshold $m_\pi\sim 140\,$MeV, and the theory 
$SU(2)_\tau$ is `contaminated' by the hadronic world and weak 
interactions: Fire-balls generated by momentum transfers
higher than $m_\pi$, which rather often occur in processes involving the $\tau$ lepton, 
do contain strongly and weakly decaying hadronic resonances. As a consequence, 
the factor $SU(2)_\tau$ should be treated separately. So far we do not understand why quarks and charged 
hadrons couple to the same photon as the electron does. It is possible 
that the three (almost) massless photons of 
SU(2)$_{\tiny\mbox{CMB}}\times$SU(2)$_{e}\times$SU(2)$_{\mu}$, 
which contribute to scattering processes at momentum transfers larger than 
$T^\mu_0$ but lower than $T^{\tiny\mbox{QCD}}_0$, are quantum entangled with very heavy 
photons originating from a pure SU(N) gauge theory disguising itself as 
QCD in its confining phase. The three massless photons would then see the same electric 
charge as the QCD photons saw at temperatures considerably 
larger than $T^{\tiny\mbox{QCD}}_0$. So far we can only speculate about 
the pure gauge theory which generates quarks. More definite investigations will 
be performed in the future. The exponentials parametrized 
by $T^e_0\sim 1/200\,T^\mu_0$ are practically 
unity for $\sqrt{q^2}$ sufficiently below $T^e_0$. As a consequence, 
the form factors of $e^{\pm}$ or $\mu^{\pm}$ are unity. Both the electron and the muon appear to be 
pointlike in elastic scattering. This statement should generalize to annihilation, pair production, 
or Compton scattering. For momentum transfer much larger than $T^e_0$ but smaller than $T^\mu_0$ 
the first exponential in (\ref{FF}) is strongly suppressed as compared to the second one. 
At first sight one would conclude that the 
effective form factor is 1/2 instead of the
measured value unity. However, the loss in form factor is {\sl precisely} compensated by the 
occurrence of an additional photon in the fire-ball (for momentum transfer considerably 
larger than $T_0^e$ the SU(2)$_e$ gauge dynamics is 
in {\sl addition} to SU(2)$_{\tiny\mbox{CMB}}$ in its {\sl electric} phase, 
see \cite{Hofmann20032}). This effectively multiplies the left-over exponential 
by a factor of two. The resulting form factor 
for $e^{\pm}$ or $\mu^{\pm}$ is, again, unity -- in perfect agreement with 
experiment \cite{Barber1953,Scott1951}. Approaching $T_0^e$ from below, 
the fall-off of the exponential is not yet 
compensated by an additional, massless photon: the cross section is below the QED prediction. 
This situation is, indeed, an experimental result \cite{Ashkin1953,Barber1953}. For 
$\sqrt{q^2}$ much larger than $T_0^e$ but lower than $T_0^\mu$ the electron should 
appear structureless again. This is verfied in an experiment \cite{Scott1951}. For momentum 
transfers $\sqrt{q^2}$ considerably larger than $T^\mu_0$ 
we run into hadronic contaminations which we are not able to handle at present. 
For $\sqrt{q^2}$ slighly higher than $T^\mu_0$ the left over exponential in (\ref{FF}) 
starts to become active. This is, again, compensated by the kinematical liberation of an 
additional photon in the fire-ball. As a consequence, we do not expect a drastic deviation from unity
of the effective form factor for $e^\pm$ as we cross the muon threshold. 
This expectation is in agreement with the 
experimental data in \cite{Alvensleben1968,Asbury1967} but not 
in \cite{Blumenthal1966}. In conclusion, we have made sufficiently clear that 
the structure of the electron and the muon 
is only resolved in 
elastic scattering experiments for momentum transfers close to the masses of 
these charged leptons. \\ 
\indent {\sl Possible explanation of missing} $\Lambda_{\tiny\mbox{cosmo}}$. Although 
topically not in the main line of this Letter we use the remaining space to 
speculate on the mechanism which generates the part in 
$\Lambda_{\tiny\mbox{cosmo}}$ not generated by the strongly interacting gauge 
theory underlying electromagnetism \cite{Hofmann20032}. 
Let us assume that all particles and their interactions are generated 
by products of SU(N) Yang-Mills theories. As a consequence, stable and massive particles are 
solitons of these theories in their confining phases. 
Particle collisions with momentum transfers larger than the associated Hagedorn 
temperature $T^{\tiny\mbox{SU(N)}}_0$ generate 
local fire-balls where the gauge theory is not in its confining 
phase. Locally, the theory generates {\sl negative} 
pressure \cite{Hofmann20031} shortly before the fire-ball dissolves. 
For example, shortly before the formation of the $e^-e^-$ pair the theories 
SU(2)$_e$ or SU(2)$_\mu$ generate 
negative pressure inside the local 
fire-ball generated in elastic $e^-e^-$ collisions at $\sqrt{s}$ higher 
than $T_0^{e}$ or $T_0^{\mu}$. A similar situation holds in 
relativistic hadron or heavy-ion collisions. Space expands quasi-exponentially 
in time within fire-balls of equation of state close to 
$\rho=-P$. This is in contrast to the power-law behavior within a 
radiation dominated fire-ball occuring in the very early stages of a collision with 
momentum transfer much larger than the Hagedorn temperature of the associated SU(N) gauge theory. 
Averaging the local equation of state over space at a given time in a gravitationally 
gauge invariant way, we thus expect a negative, global 
equation of state to be generated \cite{averexp}. This coarse-grained contribution to 
$\Lambda_{\tiny\mbox{cosmo}}$ could make up for the too small \cite{WMAP} 
homogeneous part in $\Lambda_{\tiny\mbox{cosmo}}$ originating from SU(2)$_{\tiny\mbox{CMB}}$. \\ 
{\sl Acknowledgments:} 
The author would like to thank Luis Alvarez-Gaume and Arjun Berera for very stimulating 
discussions. Useful conversations with Peter Fischer, Holger Gies, J\"org J\"ackel, 
Philippe de Forcrand, Bertold Stech, Zurab Tavartkiladze, Christof Wetterich, and 
Werner Wetzel are gratefully acknowledged. The author would like to thank 
Otto Nachtmann for his very sophisticated help in finding the relevant literature. 
The financial support and the hospitality of CERN's theory division are thankfully 
acknowledged. 

\bibliographystyle{prsty}

\begin{thebibliography}{10}
\bibitem{expl}
The onset of the magnetic phase of SU(2)$_{\tiny\mbox{CMB}}$, in which the present Universe resides, 
is determined by $T_{\tiny\mbox{CMB}}$ at which the photon $\gamma$ is precisely massless.
\bibitem{neutrinos} 
In this Letter we do not discuss a Hagedorn temperature which 
could be defined with respect to the continuous spectrum of neutrino 
excitations (center-vortex loops with varying size).
\bibitem{averexp}
The pressure outside of a {\sl local} fire-ball is that of the {\sl global} 
fire-ball generated by SU(2)$_{\tiny\mbox{CMB}}$. It is negative \cite{Hofmann20032}.
\end{thebibliography}

\end{document}